# Multiparticle tree amplitudes in scalar field theory

F. T. Brandt and J. Frenkel

*Instituto de Física, Universidade de São Paulo, São Paulo, 05389-970 SP, Brazil*

**Abstract**

Following an argument advanced by Feynman, we consider a method for obtaining the effective action which generates the sum of tree diagrams with external physical particles. This technique is applied, in the unbroken $\lambda\phi^4$ theory, to the derivation of the threshold amplitude for the production of $n$ scalar particles by $n$ initial particles. The leading contributions to the tree amplitude, which become singular in the threshold limit, exhibit a factorial growth with n.

## 1 Introduction

There has been recently much work on the problem of calculating amplitudes of multiparticle production at threshold [1–10]. Even for processes involving weakly interacting particles, multiparticle amplitudes may become important since these are associated with a large number of diagrams in perturbation theory. This situation arises for example in the $\lambda\phi^4$ model, when a virtual scalar particle produces a large number of on-mass-shell particles. In an elegant paper [11], Brown has shown that the sum of the corresponding tree graphs is generated by the solution of the classical field equation. Due to the fact that the process of particle production involves only the positive frequency part of the field, this solution is necessarily complex.

We consider here a related method, outlined by Feynman [12], for obtaining the tree amplitudes with asymptotic free particles. This technique also enables the calculation of processes involving both absorbtion and production of physical particles. Although the method can be applied to any field theory, we shall discuss for simplicity only a scalar theory involving self-interacting fields $\phi$. (For the case of a gauge theory, see also Ref. [13]).

The Lagrangian density for the system is given by

$$\mathcal{L}(\phi) = \frac{1}{2}\left(\partial_\mu \phi \partial^\mu \phi - m^2 \phi^2\right) - V(\phi), \tag{1.1}$$

where the quadratic terms describe the propagators of the fields, and $V(\phi)$ represent all the interactions. The connected tree amplitudes in the presence of external sources $s$ are generated by the effective action [14]

$$W = -\int d^4x \left[\mathcal{L}(\phi) + \phi s\right] \tag{1.2}$$

For our purpose, it is sufficient to take $s$ infinitesimal, which resolves certain ambiguities of surface integrals at infinity. Let us denote by $\phi_0$ a general superposition of asymptotic fields representing particles coming in at $t \to -\infty$ and going out at $t \to +\infty$. We suppose for the moment that these fields are not strictly free, but obey the equation of motion

$$\left(\Box^2 + m^2\right)\phi_0 = s. \tag{1.3}$$

Thus we have that

$$\int d^4x \phi s = \int d^4x \phi \left(\Box^2 + m^2\right)\phi_0 = \int d^4x \phi_0 \left(\Box^2 + m^2\right)\phi. \tag{1.4}$$

Making use of the classical equation of motion

$$\left(\Box^2 + m^2\right)\phi = -\frac{\delta V(\phi)}{\delta \phi} + s \equiv -V'(\phi) + s, \tag{1.5}$$

we can write (1.4) in a more general form as follows

$$\int d^4x \phi s = a \int d^4x \phi_0 \left(-V'(\phi) + s\right) + b \int d^4x \phi \left(\Box^2 + m^2\right)\phi_0, \tag{1.6}$$

where $a$ and $b$ are constants subject to the condition that $a + b = 1$. Substituting (1.6) into the action (1.2), we integrate by parts and take finally the limit $s \to 0$. With the help of Eq. (1.3), we obtain for the effective action with vanishing source

$$W = \int d^4x \left[\frac{1}{2}\phi_{cl}\left(\Box^2 + m^2\right)\phi_{cl} + a\phi_0 V'(\phi_{cl}) + V(\phi_{cl})\right], \tag{1.7}$$

where $\phi_{cl}$ satisfies the classical equation of motion (1.5) in the absence of sources. In order to get agreement with the lowest order perturbation theory, it is necessary that $a = 1/2$. Therefore, the effective action for tree amplitudes with asymptotically-free particles can be written as

$$W = \int \mathrm{d}^4 x \left[ \frac{1}{2} \left( \phi_0 - \phi_{cl} \right) V' \left( \phi_{cl} \right) + V \left( \phi_{cl} \right) \right]. \tag{1.8}$$

In section 2 we consider, for definiteness, an application of this method in the context of the $\lambda \phi^4$ scalar field theory. (The $\lambda \phi^3$ model is similar in many respects.) We shall study the tree amplitude for the production of $n$ scalar particles by $n$ initial particles, in the threshold limit. Since in this process the number of particles with negative frequency equals that of particles with positive frequency, the solution of the corresponding field equation may become real. This solution is expressed in closed form in terms of the Jacobian elliptic functions [15]. We also compute explicitly the effective action which generate these amplitudes. In section 3, we compare the threshold results obtained by the above techniques with the ones determined perturbatively in the non-relativistic domain, where the 3-momenta of the external particles are much smaller than their mass. We point out that in this case the individual contributions to the amplitudes may become very large, due to the fact that the internal particles can approach their on-mass-shell regime. We also discuss certain configurations where a cancellation of the most singular terms occurs when adding the contributions of the tree graphs. The result for the uncancelled leading contributions indicates a factorial growth of the tree amplitudes with the number of particles participating in the process.

## 2 The threshold amplitude $n \to n$ in the $\lambda \phi^4$ theory

Let us consider in this model the classical field equation (1.5) with vanishing source:

$$\left( \Box^2 + m^2 \right) \phi_{cl} \left( x \right) = -\frac{\lambda}{3!} \phi_{cl}^3 \left( x \right) \tag{2.1}$$

In the threshold limit, when the 3-momenta of the external particles vanish, the classical field $\phi_{cl}(x)$ becomes a spatially uniform but time-dependent function, obeying the ordinary differential equation

$$\left[ \frac{\partial^2}{\partial t^2} + m^2 \right] \phi_{cl} \left( t \right) + \frac{\lambda}{3!} \phi_{cl}^3 \left( t \right) = 0. \tag{2.2}$$

The boundary condition is that asymptotically, as $\lambda \to 0$, $\phi_{cl}(t)$ must reduce to a general superposition $\phi_0$ of incoming fields as $t \to -\infty$ and outgoing fields

as $t \to +\infty$:

$$\phi_0 = \left(\sum_j \alpha_j^-\right) e^{-imt} + \left(\sum_j \alpha_j^+\right) e^{+imt}. \tag{2.3}$$

The coefficients $\alpha_j^-$ and $\alpha_j^+$ are associated respectively with each of the initial and final particles participating in an arbitrary process. We can write $\phi_0$ in a more compact form as

$$\phi_0 = A \sin(m\,t + \theta), \tag{2.4}$$

where

$$A = 2\left(\sum_j \alpha_j^- \sum_l \alpha_l^+\right)^{1/2}; \quad \theta = \arctan\left(\frac{\sum_j \left(\alpha_j^+ + \alpha_j^-\right)}{i\sum_l \left(\alpha_l^+ - \alpha_l^-\right)}\right). \tag{2.5}$$

The proper solution of the differential equation (2.2) which reduces asymptotically to $\phi_0$, can be expressed in terms of the elliptic function sn as follows:

$$\phi_{cl}(t) = A\,\operatorname{sn}\left[\frac{mt + \theta}{\sqrt{1-k^2}}, -k^2\right], \tag{2.6}$$

where the modulus $k^2$ is given by

$$k^2 = \frac{\lambda A^2}{12\,m^2 + \lambda A^2}. \tag{2.7}$$

Expanding $\phi_{cl}(t)$ in powers of the coupling constant $\lambda$

$$\phi_{cl}(t) = \phi_0(t) + \lambda \phi_{cl}^{(1)}(t) + \cdots, \tag{2.8}$$

it can be verified that $\phi_{cl}^{(1)}(t)$ is proportional to $\phi_0^3(t)$, in accordance with the recursive relations implied perturbatively by the field equation (2.2). The above expansion requires the condition $mtk^2 \ll 1$, which we assume to hold in what follows.

We now substitute the relation (2.6) into the effective action (1.8), obtaining

$$W = \frac{\lambda}{4!} \int d^4x \, [2\phi_0(t) - \phi_{cl}(t)] \, \phi_{cl}^3(t). \tag{2.9}$$

The spatial integrals give then the 3-dimensional volume, which is usually normalized by setting it equal to one [6]. When doing the time integration, we can perform a shift so that the action becomes effectively independent of the parameter $\theta$ introduced in (2.5).

It is appropriate to comment at this point on the following feature related with the threshold limit. In general, energy conservation is ensured by the time integral

$$\lim_{\tau \to \infty} \int_{-\tau}^{\tau} dt \, e^{i \sum_j \left(\omega_j^{out} - \omega_j^{in}\right) t} = 2\pi \delta \left[ \sum_j \left( \omega_j^{out} - \omega_j^{in} \right) \right]. \tag{2.10}$$

At the threshold, all the $\omega_j$ are set from the beginning equal to $m$. Hence, we will interpret an overall factor of $2\tau$ as being proportional to Dirac's delta function associated with the conservation of energy.

In order to perform in (2.9) the time integration over the first term in the integrand which is proportional to $\phi_0$, we use the standard representation of $\text{sn}^3$ as a trigonometric series involving the sine functions [15]. After some calculation, in the region of large values of $\tau$ such that $m\tau \gg 1$, we arrive at

$$W_a = \frac{2\lambda}{4!} \int_{-\tau}^{+\tau} dt \, \phi_0(t) \phi_{cl}^3(t) \simeq \frac{\lambda}{4!} \frac{2\pi A^4}{m} \frac{(1-k^2)}{(-k^2)^{3/2} \, \text{K}(-k^2)} \frac{q^{1/2}}{q-1}$$

$$\times \left[ 1 + \frac{\pi}{2\sqrt{1-k^2} \text{K}(-k^2)} \right] \sin \left[ \left( \frac{\pi}{2\sqrt{1-k^2} \text{K}(-k^2)} - 1 \right) m\tau \right]$$

$$\tag{2.11}$$

where $q = \exp\left[-\text{K}(1+k^2)/\text{K}(-k^2)\right]$ and K represents the complete elliptic integral of the first kind.

With the help of the known integrals involving the Jacobian elliptic functions [15], we can perform in (2.9) the time integration over $\phi_{cl}^4$ to obtain

$$W_b = -\frac{\lambda}{4!} \int_{-\tau}^{+\tau} dt \, \phi_{cl}^4(t)$$

$$= \frac{\lambda}{4!} (2\tau) \frac{A^4}{3k^4} \left[ k^2 - 2 + 2\left(1 - k^2\right) \frac{\text{E}(-k^2)}{\text{K}(-k^2)} \right]$$

$$\tag{2.12}$$

where E denotes the complete elliptic integral of the second kind.

The modulus $k^2$ defined in (2.7) is an implicit function of the parameters $\alpha_j^-$ and $\alpha_l^+$, as can be seen from (2.5). Then the action $W = W_a + W_b$ depends on the parameters $\alpha_1^-, \alpha_1^+, \alpha_2^-, \alpha_2^+ \cdots$. The tree amplitude $T_n$ for the threshold production of $n$ particles by $n$ initial particles is proportional to the coefficient of the product $\alpha_1^- \alpha_1^+ \cdots \alpha_n^- \alpha_n^+$ in the expression of the effective action. Taking into consideration the remark made following Eq. (2.10) we obtain:

$$2\tau T_n = \prod_{j=1}^{n} \frac{\partial}{\partial \alpha_j^-} \frac{\partial}{\partial \alpha_j^+} W\left(k^2, \lambda, \tau\right) \bigg|_{\alpha^{\pm}=0} . \tag{2.13}$$

We can expand $\lambda W$ in a power series of $k^2$ as follows:

$$\lambda W\left(k^2, \lambda, \tau\right) = \left(k^2\right)^2 W^{(2)}(\tau) + \cdots + \left(k^2\right)^n W^{(n)}(\tau) + \cdots, \tag{2.14}$$

since in the present case $W^{(0)}$ and $W^{(1)}$ vanish.

It is now possible to perform the required differentiation in (2.13), obtaining for the threshold amplitude the expression

$$T_n = \frac{1}{2\tau\lambda} \left(\frac{\lambda}{3!}\right)^n \frac{(-1)^n 2^n (n!)^2}{m^{2n}} \left[ \binom{n-1}{1} W^{(2)} + \cdots \right.$$

$$\left. + \binom{n-1}{n-2} W^{(n-1)} + W^{(n)} \right] \tag{2.15}$$

For example, we find explicitly to lowest orders that

$$T_2 = \lambda; \qquad T_3 = -\frac{1}{4}\frac{\lambda^2}{m^2} \tag{2.16}$$

The complete expression for $T_n$ becomes increasingly cumbersome in higher orders, since the action $W$ given by (2.11) and (2.12) is a rather complicated function. For this reason, we shall restrict here to the determination of the leading contributions to $T_n$, which arise from $W_a$ for large values of $\tau$. To this end we recall that consistency with perturbation theory requires the condition: $m\tau k^2 < 1$ (see the comments following (2.8)). We therefore expand the argument of the sine function in (2.11) in powers of $m\tau k^2$, disregarding contributions like $m\tau k^4$, etc. Proceeding in this way, we obtain that the leading contributions to the threshold amplitude are given by

$$T_n^L = (-1)^{n/2} \frac{n}{2} n! \lambda \left(\frac{\lambda\tau}{4m}\right)^{n-2} \tag{2.17}$$

where $n \geq 4$ is a even integer. When $n$ is odd, the contributions to $T_n$ are subleading since $\tau$ occurs raised to a power smaller than $n - 2$.

In order to interpret this result, we turn now to an explicit evaluation in the non-relativistic limit of the Feynman diagrams appearing in perturbation theory.

## 3 Discussion

The lowest order contribution for the process $2 \to 2$ is trivial, being given just by $T_2$ in Eq. (2.16). We therefore start by considering the contributions of the Feynman diagrams shown in Fig. 1, to the amplitude $p_1 + p_2 + p_3 \to p_4 + p_5 + p_6$.

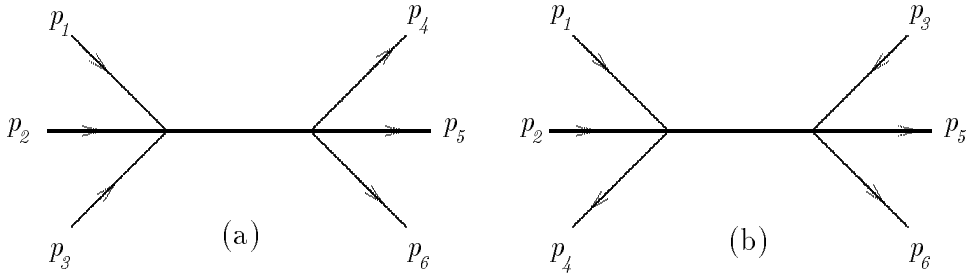

Fig. 1. Tree graphs of order $\lambda^2$. Additional permutations of the external lines in the diagram (b) are to be understood.

While the graph (1a) is regular in the threshold limit, the diagram (1b) becomes singular since the propagator approaches in this case the on-mass-shell pole. Adding the contributions of the singular diagrams we get, in the non-relativistic limit when $|\vec{p}_i| \ll m^2$ and $\omega_i \simeq m + (\vec{p}_i)^2/2m$

$$-\frac{2}{\lambda^2} \tilde{T}_3^L = \left[ \frac{1}{(\vec{p}_1 - \vec{p}_4) \cdot (\vec{p}_2 - \vec{p}_4)} + \frac{1}{(\vec{p}_2 - \vec{p}_4) \cdot (\vec{p}_3 - \vec{p}_4)} + \frac{1}{(\vec{p}_3 - \vec{p}_4) \cdot (\vec{p}_1 - \vec{p}_4)} \right] + [\vec{p}_4 \to \vec{p}_5] + [\vec{p}_4 \to \vec{p}_6] \quad (3.1)$$

The individual terms in this expression diverge when all $\vec{p}_i \to 0$. Furthermore, for fixed values of the 3-momenta, there will be additional singularities when some initial $\vec{p}_i$ become parallel and equal to some outgoing momenta $\vec{p}_{out}$.

We will consider explicitly in what follows the configuration when all $\vec{p}_i$ are parallel to a given vector $\vec{p}$: $\vec{p}_i = c_i \vec{p}$, where $c_i$ are arbitrary numbers. We find that in this case the singular contributions $\tilde{T}_3^L$ cancel out when $(\vec{p})^2 \ll m^2$, leaving only regular terms of the form indicated in (2.16). This equation followed in consequence of the classical solution (2.6). That our solution can be relevant for such a configuration may be expected, since $\phi_{cl}(t)$ represents the threshold limit $\vec{p} \to 0$ of a more general solution of the equation of motion (2.1), namely

$$\phi_{cl}(x) = A \operatorname{sn}\left[\frac{p \cdot x + \theta}{\sqrt{1-k^2}}, -k^2\right], \qquad (3.2)$$

where $p_\mu = (\omega, \vec{p})$ satisfies the on-mass-shell condition $p_\mu p^\mu = m^2$.

We consider next the Feynman diagrams of order $\lambda^3$ depicted in Fig. 2, which are singular in the threshold limit. In the non-relativistic limit, graphs like the one shown in Fig. (2a) give contributions proportional to $1/(\vec{p})^2$. On the other hand, diagrams like the one shown in Fig. 2b give, in addition to such contributions, also superleading terms proportional to $1/(\vec{p})^4$. There are altogether 216 graphs like this one and adding the contributions from all graphs we find, after a very long computation, that the superleading terms cancel out.

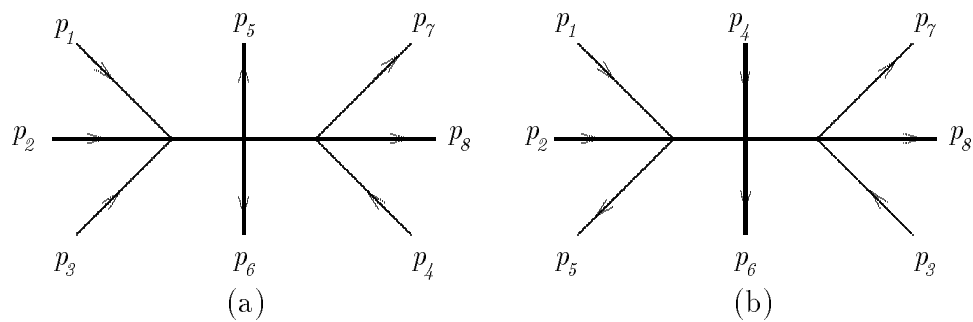

Fig. 2. Feynman diagrams contributing to the tree amplitude: $p_1 + p_2 + p_3 + p_4 \to p_5 + p_6 + p_7 + p_8$. Graphs obtained by permutations of the external lines are to be understood.

Since the general result for the leading terms of order $1/(\vec{p})^2$ is very involved, we quote here only the simple expression obtained when $\vec{p}_2 = \vec{p}_3 = \vec{p}_4 = \vec{p}_{in}$ and $\vec{p}_5 = \vec{p}_6 = \vec{p}_7 = \vec{p}_{out}$. Then, $\vec{p}_1$ and $\vec{p}_8$ are determined by energy-momentum conservation and we get

$$\tilde{T}_4^L = \frac{17}{4} \frac{1}{|\vec{p}_{in} - \vec{p}_{out}|^2} \frac{\lambda^3}{m^2} \equiv \frac{17}{4} \frac{1}{|\Delta\vec{p}|^2} \frac{\lambda^3}{m^2}. \qquad (3.3)$$

In order to compare this result with the corresponding term $T_4^L$ obtained from Eq. (2.17), we recall that in the previous analysis we have put all $\vec{p}_i = 0$, but kept $\tau$ fixed (and large). On the other hand, in ordinary perturbation theory we first let $\tau \to \infty$, and keep $\vec{p}_i \neq 0$, in order to avoid singularities. If the perturbative calculations were carried out in a space-time box of size of order $\tau$, then $\tilde{T}_4^L$ would be replaced by an expression like

$$\tilde{T}_4^L(|\Delta\vec{p}|, \tau) = \frac{17}{4} \frac{1}{|\Delta\vec{p}|^2} \left[1 - \cos(c\,\tau\,|\Delta\vec{p}|)\right] \frac{\lambda^3}{m^2}, \tag{3.4}$$

where $c$ is a constant of order unity. Clearly, as $\tau \to \infty$ the cosine function oscillates rapidly and (3.4) reduces to the previous expression (3.3). On the other hand, if we let $\vec{p}_i \to 0$ for fixed $\tau$, then $\tilde{T}_4^L$ yields a result of the form indicated by Eq. (2.17). Hence, our previous expression for the threshold amplitude $T_n^L$ can be interpreted in terms of the perturbative results in the non-relativistic limit, provided we make the correspondence $\tau \to constant/|\Delta\vec{p}|$.

Thus, we conclude that the leading contributions to the tree amplitude $n \to n$, in the configuration when the momenta are parallel and non-relativistic, have the form

$$T_n^L \sim (-1)^{n/2}\ n\ n!\ \lambda \left(\frac{\lambda}{m\,|\vec{p}_{in} - \vec{p}_{out}|}\right)^{n-2} \tag{3.5}$$

where $\vec{p}_{in}$ and $\vec{p}_{out}$ denote respectively typical momenta of the incoming and outgoing particles. Such contributions occur only when $n \geq 4$ is an even integer. For odd values of $n$, the factor $1/|\Delta\vec{p}|$ appears raised to a power less than $n - 2$, so that these contributions to $T_n$ are subleading.

The superleading terms involve large factors like $(1/|\Delta\vec{p}|)^N$, where $n - 2 < N \leq 2(n-2)$. Their cancellation emerges in perturbation theory only after long calculations and looks rather mysterious. On the other hand, the absence of such terms in the complete amplitude is naturally predicted by the expression (3.5). We remark finally the factorial growth of the tree amplitude with the number of particles involved in the scattering process. This result tends to confirm the conclusion of previous works, indicating that the perturbation theory may not converge for a sufficiently large number of particles.

**Acknowledgement**

We would like to thank CNPq (Brasil) for a grant. J. F. is grateful to Prof. J. C. Taylor for a very helpful correspondence.